\documentclass{jfm}
\usepackage{graphicx}
\usepackage{epstopdf, epsfig}
\usepackage{xcolor}
\newtheorem{lemma}{Lemma}
\newtheorem{corollary}{Corollary}

\usepackage{url} 

\newcommand{\mib}[1]{\mbox{\boldmath$#1$}}
\newcommand{\tx}[1]{\textrm{\tiny #1}}

\newcommand{\revone}[1]{\textcolor{black}{#1}}
\newcommand{\revtwo}[1]{\textcolor{black}{#1}}
\newcommand{\revthree}[1]{\textcolor{black}{#1}}

\shorttitle{Magnetostrophic Rossby wave soliton}
\shortauthor{K. Hori, S. M. Tobias, and C. A. Jones}

\title{Solitary magnetostrophic Rossby waves in spherical shells}

\author{K. Hori\aff{1,2}
  \corresp{\email{amtkh@leeds.ac.uk}},
  S. M. Tobias\aff{2}
 \and C. A. Jones\aff{2}}

\affiliation{\aff{1}Graduate School of System Informatics, Kobe University,
 1-1 Rokko-dai, Nada, Kobe, Japan %657-8501, Japan
\aff{2}Department of Applied Mathematics, University of Leeds,
 Woodhouse Lane, Leeds, UK} %LS2 9JT, UK}

\begin{document}

\maketitle

\begin{abstract}
%%This file contains instructions for authors planning to submit a paper to the {\it Journal of Fluid Mechanics}. These instructions were generated in {\LaTeX} using the JFM style, so the {\LaTeX} source file can be used as a template for submissions. The present paragraph appears in the \verb}abstract} environment. All papers should feature a single-paragraph abstract of no more than 250 words, which provides a summary of the main aims and results. 
%%
%%
%**250 words**
Finite-amplitude hydromagnetic Rossby waves in the magnetostrophic regime are studied.
We consider the slow mode, which travels in the opposite direction to the hydrodynamic or fast mode,
 in the presence of a toroidal magnetic field and zonal flow
 by means of quasi-geostrophic models for thick spherical shells.
The weakly-nonlinear, long waves are derived asymptotically
 using a reductive perturbation method.
The problem at the first order is found to obey
 a second-order ODE, leading to a hypergeometric equation for a Malkus field 
 and a confluent Heun equation for an electrical-wire field,
 and is nonsingular when the wave speed approaches the mean flow.
Investigating its neutral, nonsingular eigensolutions for different basic states,
 we find the evolution is described by the Korteweg-de Vries equation.
This implies that the nonlinear slow wave forms solitons and solitary waves. 
These may take the form of a coherent eddy,
 such as a single anticyclone.
We speculate on the relation of the anti-cyclone to the asymmetric gyre seen in Earth's fluid core,
 and in state-of-the-art  dynamo DNS.
\end{abstract}

\begin{keywords}
%**Authors should not enter keywords on the manuscript, as these must be chosen by the author during the online submission process and will then be added during the typesetting process (see http://journals.cambridge.org/data/\linebreak[3]relatedlink/jfm-\linebreak[3]keywords.pdf for the full list)
\end{keywords}

\vspace{-10mm}
\section{Introduction}  \label{sec:intro}

Linear waves in an inviscid, perfectly-conducting fluid
 permeated by a uniform magnetic field $\mib{B}_0$
 in a frame rotating with rate $\mib{\Omega}$ satisfy the dispersion relation \citep{L54}
\begin{equation}
 \omega = \pm \frac{\mib{\Omega}\cdot\mib{k}
                \pm \sqrt{(\mib{\Omega}\cdot\mib{k})^2
                          + |\mib{k}|^2 (\mib{B}_0 \cdot\mib{k})^2/\rho\mu_0} }
               {|\mib{k}|}  \; ,   \label{eq:MC_dispersion_relation_general}
\end{equation}
where $\omega$ is the frequency, $\mib{k}$ is the wavenumber vector,
 $\rho$ is the density, and $\mu_0$ the magnetic permeability.
This yields a wide variety of magnetic Coriolis (MC) waves,
 including fast (modified inertial) and slow (magnetostrophic) waves; the latter being 
unique to rotating magnetohydrodynamics (MHD).
In this manuscript
 we consider magnetostrophic waves 
 for which $(\mib{\Omega}\cdot\mib{k})^2/|\mib{k}|^2 \gg (\mib{B}_0\cdot\mib{k})^2/(\rho\mu_0)$:
In particular, one class which has the relation 
\begin{equation}
 \omega \approx - \frac{ (\mib{B}_0 \cdot \mib{k})^2 |\mib{k}|^2 }{ \rho\mu_0\beta k } \; .
   \label{eq:slowMR}
\end{equation}
Here $\beta$ denotes the beta parameter, $k$ is the azimuthal wavenumber, and the minus sign
 indicates waves travel  opposite to the hydrodynamic Rossby wave, $\omega = \beta k/|\mib{k}|^2$. 
This class is sometimes referred to as slow hydromagnetic-planetary or magnetic-Rossby (MR) waves \citep{H66}. 
Relation (\ref{eq:slowMR}) indicates they are dispersive, and depend on the background field and the wavelength;
 these waves have been suggested to be important
 in Earth's fluid core and for the geomagnetic westward drift \citep[e.g.][]{H66,M67,CFF14,HJT15,NSKHH20}.

Other classes of MC waves include torsional Alfv\'{e}n waves
 for which $\mib{\Omega}\cdot\mib{k} \approx 0$
 and $(\mib{\Omega}\cdot\mib{k})^2/|\mib{k}|^2  \ll (\mib{B}_0\cdot\mib{k})^2/(\rho\mu_0)$
 \citep{Bra70,RA12,GJF15}. 
\revthree{More recently inertial-Alfv\'{e}n waves \citep{BD16} have been claimed
 to account for the geomagnetic jerks \citep{AF19}.}
Laboratory experiments have identified several types of magnetostrophic waves
in spherical Couette flows with a dipolar magnetic field being applied \citep{SAetal08}.
We note the wave dynamics relies on both the direction and the morphology
 of the background magnetic field,
 as illustrated in the simple planar model (\ref{eq:slowMR}).
Here we focus on the problem with a purely azimuthal basic field;
 for this case (\ref{eq:slowMR}) reduces to $\omega \propto k |\mib{k}|^2$\revthree{,
 indicating its linear and cubic relationship to the azimuthal wavenumber.}

The linear theory for MC waves in stably-stratified, thin layers is well-studied
 \citep[e.g.][]{Bra67,G00,ZOBS07,MJT17}
 as observational exploration of the geomagnetic field and the solar corona has
 developed to reveal periodic patterns \citep{CAM15,MCML17}. 
Stratification in general introduces a correction term to
 the dispersion relations of MC waves, 
 whilst in a thin layer the direction of
 travel is usually reversed; 
 however, this is not always true in spherical geometries.
The unstratified thick shell problem considered here
 is sufficient to provide some fundamental
 understanding of the nonlinear problem.

Theoretical investigation is expanding
 to consider their nonlinear properties such as 
 turbulence \citep{TDH07} and 
 triadic resonances \citep{RR15}. 
\citet{L17} found a couple of cases in which
 nonlinear equatorial waves in the shallow water MHD should be governed
 by Korteweg-de Vries (KdV) equations and so behave like solitary waves.
They were mostly fast MR modes, recovering the equatorial Rossby wave soliton \citep{Boy80}
 in the nonmagnetic limit,
 but he reported one case in which the wave would slowly travel in the opposite azimuthal direction.
\citet{H19} investigated magnetostrophic MR waves in a Cartesian quasi-geostrophic (QG) model.
The slow, weakly-nonlinear waves
  led to evolution obeying the KdV equation unless
 the basic state -- all the magnetic field, topography, and zonal flow -- is uniform.
Slow MR waves have been seen in spherical dynamo DNS  
 travelling with crests/troughs that were isolated and sharp,
 unlike the continuous wave trains
 that might be expected \citep{HJT15,HTJ18}.

Hydrodynamic Rossby wave solitons have been extensively studied,
motivated by atmosphere and ocean dynamics \citep[e.g.][]{C71,R77,Boy80}. %,L65,M82}.
In the long wave limit it has been demonstrated that the QG soliton relies on the presence of a shear in the basic flow or  topography. 
\citet{R77} further analysed nonlinear critical layers arising from singularities
 as the wave speed approaches the basic flow speed,
 and discussed their relevance for the persistence of Jupiter's Great Red Spot.

The present manuscript demonstrates that weakly nonlinear slow MR waves in spherical containers  
 yield soliton solutions. 
We adopt simple QG MHD models 
 and asymptotically derive the evolution equation for the long wave 
 when the basic magnetic field and flow are both azimuthal.
We demonstrate: 
 (i) the amplitude at the first order is described by the KdV equation
 for the chosen basic states, 
 (ii) the problem is dictated by an ODE, which 
 has no singularities as the wave speed approaches the basic flow speed,
 and (iii) the single soliton (solitary wave) solution to the KdV equation
 implies an isolated eddy that  progresses in a stable permanent form
 on magnetostrophic timescales.

\section{Theoretical foundations}

We consider an inviscid, incompressible, ideal quasi-geostrophic (QG) model of electrically conducting fluid
 within a rapidly rotating shell,
 bounded by inner and outer spheres of radii $r_\tx{i}$ and $r_\tx{o}$, respectively 
 \citep[e.g.][]{Bus70,GJ06}.
We use polar coordinates $(s,\varphi,z)$ with rotation  $\Omega \hat{\mib{z}}$.

For rapid rotation, the incompressible horizontal QG fluid motion can be expressed
 as $\mib{u} \approx \nabla \times \psi (s, \varphi) \hat{\mib{z}}$
 with $\psi$ a streamfunction, 
so it is independent of $z$.
When the magnetic field is not too strong to violate the QG approximation,
 we further assume the magnetic field may be written as
 $\mib{B} \approx \nabla \times g (s, \varphi) \hat{\mib{z}}$
 with $g$ being the potential
 \citep[e.g.][]{Bus76,AJPJ00,TDH07,CFF14}.
 No penetration on 
 the spherical boundaries
 at $z=\pm H = \pm \sqrt{r_\tx{o}^2 - s^2}$
 enables us to represent the Coriolis term of the axial vorticity equation 
 in terms of the topography-induced beta parameter. The equations for the $z$-components of the vorticity
 and the magnetic potential in dimensionless form are then:
\begin{eqnarray}
 \frac{\partial}{\partial t} \Delta_\tx{H} \psi 
  - \mathcal{J} [ \psi, \Delta_\tx{H} \psi ]
  - \frac{1}{Le^2} 
     \frac{\beta}{s} \frac{\partial \psi}{\partial \varphi}
  &=& - \frac{1}{Le^2} 
        \mathcal{J} [ g, \Delta_\tx{H} g ]  \\
\mbox{and\ } \quad
  \frac{\partial}{\partial t}  g
   &=&  \mathcal{J} [ \psi, g ]      \; ,   \label{eq:current_sphere}
\end{eqnarray}
where  $\Delta_\tx{H} = (1/s) \partial/\partial s (s \partial/\partial s) + ( 1/s^2 ) \partial^2/\partial \varphi^2$, 
and 
$\mathcal{J} [f_1,f_2]
 =  ( \partial f_1/\partial s \; \partial f_2/\partial \varphi
     -\partial f_2/\partial s \; \partial f_1/\partial \varphi )/s$
for any functions $f_1$ and $f_2$. 
Here the length, the magnetic field, and the velocity
 are, respectively, scaled by the radius of the outer sphere $r_\tx{o}$,
 the mean field strength $B_0$
 and the MC wave speed
 $B_0^2/(2\Omega r_\tx{o} \rho \mu_0) = c_\tx{M}^2/c_\tx{C}$; %= c_\tx{MC}
 $c_\tx{M}^2 = B_0^2/(\rho\mu_0)$ and $c_\tx{C} = 2\Omega r_\tx{o}$.
The Lehnert number $Le %= (c_\tx{MC}/c_\tx{C})^{1/2}
 = c_\tx{M}/c_\tx{C}$, whilst the beta parameter is given by $\beta = s/(1 - s^2)$. 
Impermeable boundary conditions are applied so that 
\begin{eqnarray}
 \frac{1}{s} \frac{\partial \psi}{\partial \varphi}= 0 
   \quad &\mbox{at\ }& \quad s =  \eta, 1 ,
  \label{eq:bc_sphere}
\end{eqnarray}
where the aspect ratio $\eta = r_\tx{i}/r_\tx{o}$.
As $\beta \to \infty$ at $s=1$, the governing equations are singular there; these boundary conditions ensure that the regular solution is selected.

Of particular interest is the regime when $Le^{-1}$ is large.  
Taking the limit leads to a balance between the vortex stretching and the Lorentz term in the vorticity equation:
\begin{eqnarray}
  \beta \frac{1}{s} \frac{\partial \psi}{\partial \varphi}
  &=& \mathcal{J} [ g, \Delta_\tx{H} g ] 
       \; ,  \label{eq:vorticity_sphere_magnetostrophic}
\end{eqnarray}
whilst (\ref{eq:current_sphere}) retains its same form.
%%
%The linear problem yields a dispersion relation revealing the retrograde propagation (see below). 
%
%The linear problem comprising (\ref{eq:vorticity_sphere_magnetostrophic}), (\ref{eq:current_sphere}), and (\ref{eq:bc_sphere})
 %may be treated analytically for special cases.
%For instance, when the basic magnetic field increases linearly with $s$,
% $\beta$ is uniform, and basic flows are absent, 
 %the linearised equations are reduced to Bessel's differential equation \citep{CFF14}
 %and yield $\omega = - \eta_n^2 m/4$ %, i.e. retrograde propagation.
 %where $\eta_n$ is the $n$-th eigenvalue for the azimuthal wavenumber $m$. 
%
The nonlinear problems have two source terms acting on the magnetostrophic wave: 
below we asymptotically solve the weakly nonlinear cases.

To seek solitary long-wave solutions
we introduce slow variables with a small parameter $\epsilon$ ($\ll 1)$ and a real constant $c$: 
\begin{equation}
 \tau = \epsilon^{3/2} t \;, \qquad
 \zeta = \epsilon^{1/2} (\varphi - c t) \; .
\end{equation} 
Note that this assumes a long spatial scale in the azimuthal direction compared with the radial direction. This is reasonable for small $m$. We then expand variables with $\epsilon$ as 
\begin{eqnarray}
 \psi = \psi_0 (s) + \epsilon \psi_1 (s,\zeta,\tau)
                   %+ \epsilon^2 \psi_2 (s,\zeta,\tau)
		   + ...   \; , \quad 
    g =    g_0 (s) + \epsilon g_1 (s,\zeta,\tau)
                   %+ \epsilon^2 g_2 (s,\zeta,\tau)
                   + ...  \;,  
\end{eqnarray}
for the  basic state satisfying
\begin{equation}
  - D \psi_0
   = \overline{U}(s)    \; , \quad 
  - D g_0
   = \overline{B}(s), 
\end{equation}
where $D = d/ds$.
At zeroth order the equations of vorticity (\ref{eq:vorticity_sphere_magnetostrophic})
 and of electric potential (\ref{eq:current_sphere}), and the boundary condition (\ref{eq:bc_sphere}) are all trivial.
%; the boundary condition (\ref{eq:bc_sphere}) leaves 
%\begin{equation}
% D^2 g_0 
%  = \frac{1}{s} D s \overline{B}
%  = 0 
%  \quad\mbox{at\ }\quad  s = \eta, 1  
% \label{eq:g0_bc_sphere}
%\end{equation}
%where $D^2 = (1/s) D s D$.

%
%
At $\mathcal{O}(\epsilon)$,
 (\ref{eq:vorticity_sphere_magnetostrophic}) and (\ref{eq:current_sphere}) become 
\begin{equation}
  \beta  \frac{\partial \psi_1}{\partial \zeta}
  = - \left[  \overline{B} \mathcal{D}^2  - \overline{J} \right] \frac{\partial g_1}{\partial \zeta},
   \qquad \textrm{where} \qquad
  \mathcal{D}^2 = \frac{1}{s} \frac{\partial}{\partial s} s \frac{\partial}{\partial s} \quad \textrm{and} \quad
  \overline{J} = D \frac{1}{s} D (s \overline{B}) 
   \label{eq:psi1_sphere} 
\end{equation}
% $\overline{J} = D ((1/s) D (s \overline{B}))$, and 
\begin{equation}
 %\overline{J} = D ((1/s) D (s \overline{B})); \qquad \qquad
 \textrm{and} \qquad
  \left( \frac{\overline{U}}{s}  - c  \right) 
     \frac{\partial g_1}{\partial \zeta}
    = \frac{\overline{B}}{s} 
     \frac{\partial \psi_1}{\partial \zeta}    
     \; ,   \label{eq:g1_sphere} 
\end{equation}
respectively.
Substituting (\ref{eq:psi1_sphere}) into (\ref{eq:g1_sphere}) gives
 a homogeneous PDE with respect to $g_1$:
\begin{equation}
 \mathcal{L} \frac{\partial g_1}{\partial \zeta} \equiv 
   \left\{ 
            \frac{\overline{B}}{ \beta s} \left[ \overline{B} \mathcal{D}^2   
                       - \overline{J} \right]  
   +  \left( \frac{\overline{U} }{s} - c \right) 
   \right\} \frac{\partial g_1}{\partial \zeta} 
 = 0 
   \label{eq:g1_pde_sphere}
\end{equation}
where $\mathcal{L}$ represents the linear differential operator
 comprising of $s, {\partial/\partial s}$ or $D, \overline{B},\beta, \overline{U}$, and $c$. 
Inserting the boundary conditions (\ref{eq:bc_sphere}) at this order into (\ref{eq:g1_sphere}) yields
\begin{eqnarray}
  \frac{\partial g_1}{\partial \zeta} 
  = 0 
   \quad \mbox{at\ } \quad s = \eta , 1 \; .
\end{eqnarray}
We then seek a solution in the form of $g_1 = \Phi(s) G(\zeta,\tau)$, so that 
\begin{equation}
 \mathcal{L} \Phi  = 0 
 \qquad \mbox{and\ } \qquad
   \Phi =  0  \quad\mbox{at\ }\quad  s = \eta, 1 
   \; . 
 \label{eq:g1_ode_sphere}
\end{equation}
Now the linear operator $\mathcal{L}$ is the ordinary differential operator
 with the partial derivatives with respect to $s$ replaced by $D$. 
Given a basic state, 
 the ODE (\ref{eq:g1_ode_sphere}) together with the boundary conditions is 
 an eigenvalue problem to determine the eigenfunction $\Phi$ with eigenvalue $c$; 
 it can have many eigensolutions.  
We note that the second-order ODE (\ref{eq:g1_ode_sphere}) remains non-singular
 as $\overline{U}/s \rightarrow c$, 
 but not as $\overline{B}^2/\beta \rightarrow 0$ unless $s = 0$.
Below we concentrate on cases in which (\ref{eq:g1_ode_sphere}) has no internal singularities,
 i.e. there is a discrete spectrum.  
We consider cases where the $z$-averaged 
 toroidal magnetic fields do not pass through zero %\citep[e.g.][]{SJNF17,HTJ18}.
 \revthree{(e.g. figure 3 of \citet{SJNF17}; figures 1-2 of \citet{HTJ18})}.
%Quasi-stable dipolar solutions in spherical dynamo DNS tend to yield the $z$-mean,
% toroidal magnetic fields do not cross zero \citep[e.g.][]{SJNF17,HTJ18}.

%
%
We proceed to the next order to obtain the amplitude function. 
Eqs.~(\ref{eq:vorticity_sphere_magnetostrophic}) and (\ref{eq:current_sphere}) at $\mathcal{O}(\epsilon^2)$ 
yield
\begin{equation}
  \beta  \frac{\partial \psi_2}{\partial \zeta}
  = - \left[ \overline{B} \mathcal{D}^2  - \overline{J}
      \right]  \frac{\partial g_2}{\partial \zeta} 
   - \frac{\overline{B}}{s^2} \frac{\partial^3 g_1}{\partial \zeta^3}  
   + \left(  \frac{\partial g_1}{\partial s} \frac{\partial}{\partial \zeta}
		         - \frac{\partial g_1}{\partial \zeta} \frac{\partial}{\partial s}
		    \right) \mathcal{D}^2 g_1
     \label{eq:psi2_sphere}
\end{equation}
\begin{equation}
 \textrm{and} \qquad
 \left( \frac{\overline{U}}{s} - c \right)  \frac{\partial g_2}{\partial \zeta}
 -  \frac{\overline{B}}{s}  \frac{\partial \psi_2}{\partial \zeta}
 =   - \frac{\partial g_1}{\partial \tau}
     + \frac{1}{s} \left(  \frac{\partial \psi_1}{\partial s} \frac{\partial g_1}{\partial \zeta}
		         - \frac{\partial \psi_1}{\partial \zeta}\frac{\partial g_1}{\partial s}
		   \right) 
          \; .  \qquad  \label{eq:g2_sphere}
\end{equation}
Eliminating $\psi_2$ using  (\ref{eq:psi2_sphere}) and $\psi_1$ using (\ref{eq:psi1_sphere}), 
 (\ref{eq:g2_sphere}) becomes the inhomogeneous PDE 
\begin{eqnarray}
 \mathcal{L} \frac{\partial g_2}{\partial \zeta}
 &=& - \frac{\overline{B}^2}{s^3 \beta} \frac{\partial^3 G}{\partial \zeta^3} 
 \Phi  
  - \frac{\partial G}{\partial \tau}  \Phi  \nonumber \\  
 &+& G \frac{\partial G}{\partial \zeta} \left\{  \frac{2\overline{B}}{\beta s} 
       \left[ (D \Phi) D^2 \Phi  -  \Phi  D (D^2 \Phi)  \right]
	     - \frac{\Phi D^2 \Phi}{s} D \left(\frac{\overline{B}}{\beta} \right)
             + \frac{\Phi^2}{s} D \left(\frac{\overline{J}}{\beta} \right) 
       \right\}
   \; %.
  \quad \qquad   \label{eq:g2_pde_sphere} 
\end{eqnarray}
where $D^2 = (1/s) D s D$. The boundary conditions here are 
\begin{equation}
 \frac{\partial g_2}{\partial \zeta}  = 0 
   \quad \mbox{at\ } \quad s =  \eta , 1 \; . 
\end{equation}
The adjoint linear problem corresponding to (\ref{eq:g1_pde_sphere}) is
\begin{equation}
\mathcal{L}^\dag \Phi^\dag  \equiv 
 \left\{ \left[ 
  D^2 \overline{B} 
 - \overline{J} \right] \frac{\overline{B}}{\beta s } 
   + \left( \frac{\overline{U}}{s} - c \right) \right\} \Phi^\dag = 0  
 \; .  \label{eq:g1_ode_adjoint_sphere}
\end{equation}
The adjoint boundary conditions are 
\begin{equation}
 \frac{\overline{B}^2}{s\beta} \Phi^\dag  = 0  
    \quad \mbox{at\ } \quad s =  \eta , 1 \; . 
 \label{eq:g1_bc_adjoint_sphere}
\end{equation} 
Note that the substitution $ \overline{B}^2 \Phi^\dag / s \beta = \Phi$ reduces the adjoint problem to the 
ordinary linear problem  (\ref{eq:g1_pde_sphere}) so, provided $\overline{B}^2 \Phi^\dag /s\beta$ is non-zero 
in the sphere, the adjoint eigenfunction $\Phi^\dag$ can simply be found by dividing the solution of
 (\ref{eq:g1_pde_sphere})  %by $\overline{B}^2 \Phi^\dag / s \beta$.}
 by $\overline{B}^2/s\beta$.

The solvability condition to (\ref{eq:g2_pde_sphere}) is thus given by
\begin{equation}
 \frac{\partial G}{\partial \tau}
 + \alpha \; G \frac{\partial G}{\partial \zeta}
 + \gamma \; \frac{\partial^3 G}{\partial \zeta^3} = 0,   
   \label{eq:KdV}
\end{equation}
 where $\alpha = \alpha_0/\delta_0$, $\gamma = \gamma_0/\delta_0$, 
\begin{eqnarray}
&&\alpha_0 =  \int _{\eta}^1 \Phi^\dag \left\{  \frac{2\overline{B}}{\beta} 
       \left[ \Phi  D (D^2 \Phi)  - (D \Phi) D^2 \Phi \right]
	     + \Phi (D^2 \Phi) D \left(\frac{\overline{B}}{\beta} \right) 
             - \Phi^2 D \left(\frac{\overline{J}}{\beta} \right)
       \right\}  \,ds
   ,  \nonumber \\
&&\gamma_0 = 
 \int_{\eta}^{1}  \Phi^\dag \frac{\overline{B}^2}{s^2 \beta}  \Phi 
   \, ds
 , %\;, %\nonumber 
  \quad \mbox{and\ } \quad
  \delta_0  = \int_{\eta}^{1} {\Phi^\dag \Phi} \  s \, ds  . \qquad \quad 
    \label{eq:g2_KdV_sphere} 
\end{eqnarray} 
Eq.~(\ref{eq:KdV}) is the Korteweg-de Vries equation
 if the coefficients, $\alpha$ and $\gamma$, are both nonzero.
In the following section we examine the coefficients
 for different choices of the basic state.

We note that the presence of $\overline{U}$ does not directly impact either $\alpha$ or $\gamma$.
It however dictates $\Phi$ and $\Phi^\dag$ through the linear problems
 at $\mathcal{O}(\epsilon)$
 and then may contribute to the terms at $\mathcal{O}(\epsilon^2)$.
This is in contrast with the hydrodynamic case \citep[e.g.][]{R77},
where the basic flow enters the nonlinear term at $\mathcal{O}(\epsilon^2)$ too.
The mean-flow effect on the magnetostrophic wave arises from the equation for the
 magnetic potential (\ref{eq:current_sphere}).

Solutions to (\ref{eq:KdV}) may take the form of solitary (single or multiple soliton),
 cnoidal, similarity, and rational waves \citep[e.g.][]{W74,DJ89}. 
For instance, for a single soliton  the asymptotic solution up to $\mathcal{O}(\epsilon)$ is  
\begin{eqnarray}
 g (s,\varphi, t)
 &&%= g_0 + \epsilon g_1
  = -\int_{\eta}^{s} \overline{B} ds + \epsilon \; \mathrm{sgn}(\alpha \gamma) \; 
             \Phi \;{\mathrm{sech}^2 F},
     \label{eq:single-soliton-g_sphere} \\  
 \psi (s,\varphi, t)
 &&%= \psi_0 + \epsilon \psi_1      
  = -\int_{\eta}^{s} \overline{U} ds
          - \epsilon \;  \mathrm{sgn}(\alpha \gamma) \;
              \left( \frac{\overline{B}}{\beta} D^2 \Phi - \frac{\overline{J}}{\beta} \Phi  \right) \;
	      {\mathrm{sech}^2 F}, 
       \label{eq:single-soliton-psi_sphere}
\end{eqnarray}
where
\begin{equation}
F (\varphi, t) 
%   = \kh{\sqrt{ \left| \frac{\alpha}{12 \gamma} \right|}
%            \left[ 
%	       \zeta - \mathrm{sgn} (\gamma) \frac{|\alpha|}{3} \tau 
%	    \right] }
   = \sqrt{  \frac{\alpha }{12 \gamma} \mathrm{sgn}(\alpha \gamma) }
	    \left[  \epsilon^{1/2} (\varphi -ct)
	          - \epsilon^{3/2} \; \mathrm{sgn}(\alpha \gamma) 
                   \; \frac{\alpha t}{3}   
	    \right] \; .
\end{equation}
This is an eddy that has the solitary characteristics in azimuth,
 riding on the basic state with the linear wave speed. 
The finite-amplitude effect $\alpha$ accelerates the retrograde propagation
 if $\gamma < 0$, but decelerates it when $\gamma > 0$.
The characteristic waveform is clearly visible in the magnetic potential.

\section{Illustrative examples}

We solve the eigenvalue problem (\ref{eq:g1_ode_sphere})
 and the adjoint problem (\ref{eq:g1_ode_adjoint_sphere})-(\ref{eq:g1_bc_adjoint_sphere})
 for different basic states
 and calculate the respective coefficients of the evolution equation (\ref{eq:KdV})
 in a spherical cavity, 
 %where $\beta = s/(1-s^2)$ 
 with $\eta = 0.35$.
%\revthree{To illustrate the connection from linear theory,}
We consider \revthree{three cases investigated in \citet{CFF14}};  
 the first has a $\overline{B}$ that is a linearly increasing function of $s$
 (referred to as a Malkus field hereafter),
 %whilst for
 the second $\overline{B}$ is inversely proportional to $s$ (an electrical-wire field)\revthree{,
 and the third one is $(3/2) \cos{\{ \pi(3/2 - 50 s/19) \} } + 2$, 
 which was adoped by \citet{CFF14} to model a profile of the radial magnetic field $B_s$ within Earth's core
 (a CFF field)}.
For %both
 \revthree{the Malkus and wire}
 fields the terms $\overline{J}$  %$D (1/s) D s\overline{B}$
 in (\ref{eq:g1_ode_sphere}), (\ref{eq:g1_ode_adjoint_sphere}) and (\ref{eq:g2_KdV_sphere})
 all vanish\revthree{, whereas this is not the case for the CFF field}.
The Malkus field
 case has been extensively studied in the literature \citep[e.g.][]{M67,RL79,ZLS03,MJT17}.
We also consider the inclusion of a basic zonal flow $\overline{U}$
 that is prograde with either a linear or quadratic dependence on $s$.

Table \ref{table:cases_spheres} summarises the  results,
 listing the eigenvalue $\lambda = \sqrt{|c|}/2$ (see below) and $c$ for the $n$-th mode,
 the coefficients $\alpha$, $\gamma$, and $\delta_0$ as calculated from the eigenfunction $\Phi$,
 the adjoint eigensolution $\Phi^\dag$ and (\ref{eq:g2_KdV_sphere}),
 and whether/at which $s$ the wave speed $c$ approaches the basic angular velocity $\overline{U}/s$.
Here the $n$-th mode has $(n-1)$ zeros within the explored interval.
Negative values of $c$ indicate retrograde waves.
More notably, in the all cases we obtain
 nonzero $\alpha$ and $\gamma$ for all $n$ examined and so the KdV equations are appropriate.
\revtwo{The fraction $|\alpha/\gamma|$ and their signs characterise the solitons.}

%%%%% NEW (ver.3.1) %%%%%%%%%%%%%%%%%%%%%%%%%%
\begin{table}
  \begin{center}
\def~{\hphantom{0}}
\begin{tabular}{ccc cccccc}
 $\overline{B}$  & $\overline{U}$ & $n$ & $\lambda$ & $c$ & $\alpha$ & $\gamma$ & $\delta_0 \times 10^{2}$ & $s$ at which $c= \overline{U}/s$ \\[3pt]   % \qquad given a normalisation factor dPhi/ds at s1
 $s$ & 0 & 1 & ~1.56402 & ~~-9.7847 & -12.854~ & 0.87465 & 4.9020~~ & --- \\
     &   & 2 & ~2.88117 & ~-33.2045 & -14.639~ & 1.0480~ & 0.79920~ & --- \\
     &   & 3 & ~4.18526 & ~-70.0655 & -26.422~ & 1.1156~ & 0.26204~ & --- \\
$1/s$& 0 & 1 & ~2.34412 & ~-21.9795 & -36.930~ & 1.2464~ & 0.92993~ & --- \\
     &   & 2 & ~4.41698 & ~-78.0389 & -31.920~ & 2.1442~ & 0.14054~ & --- \\
     &   & 3 & ~6.47665 & -167.788~ & -70.056~ & 2.8417~ & 0.044739 & --- \\
%%%%%%
 $^{\circ}s$ &  0 &  1 &   & ~~-9.7847 & -12.854~ & 0.87465 & 4.9023~~ & --- \\ % \qquad ~1.59617 & Phidag = 1 at s2
%            &    &  2 &   & ~-33.2046 & -14.638~ & 1.0481~ & 0.79935~ & --- \\ % \qquad -1.16815 & Phidag = 1 at s2
%            &    &  3 &   & ~-70.0655 & -26.412~ & 1.1157~ & 0.26211~ & --- \\ % \qquad ~0.96658 & Phidag = 1 at s2
 $^\circ 1/s$&  0 &  1 &   & ~-21.9795 & -36.865~ & 1.2464~ & 0.92800~ & --- \\ % \qquad ~0.63844 & Phidag = 1 at s2
\revthree{
 $^\circ$CFF}
            &  0 &  1 &   & ~-11.0427 & -11.493~ & 2.8531~ & 0.51035~ & --- \\ % \qquad ~1.59617 & Phidag = 1 at s2
            &    &  2 &   & ~-32.2790 & -19.611~ & 4.7250~ & 0.12427~ & --- \\ % \qquad -1.16815 & Phidag = 1 at s2
            &    &  3 &   & ~-71.6553 & -43.375~ & 4.4968~ & 0.053649 & --- \\ % \qquad ~0.96658 & Phidag = 1 at s2
%%%%%%%%%%%
 $^{\circ}s$ &$s$ &  1 &   & ~~-8.7847 & -12.854~ & 0.87465 & 4.9023~~ & none \\ % \qquad ~1.59617 & Phidag = 1 at s2 
 $^\circ 1/s$&$s$ &  1 &   & ~-20.9795 & -36.865~ & 1.2464~ & 0.92800~ & none \\ % \qquad ~0.63844 & Phidag = 1 at s2
\revthree{
 $^\circ$CFF}
             &$s$ &  1 &   & ~-10.0427 & -11.493~ & 2.8531~ & 0.51035~ & none \\ % \qquad ~1.59617 & Phidag = 1 at s2
%%%%%%%%%%%
 $^{\circ}s$ &$4s(1-s)$
                  &  1 &   & ~~-8.8379 & ~-9.5075 & 0.90339 & 4.9193~~ & none \\ % 1.59617  & Phidag = 1 at s2
 $^\circ 1/s$&$4s(1-s)$
                  &  1 &   & ~-21.4523 & -35.429~ & 1.2659~ & 0.92748~ & none \\ % 0.63844  & Phidag = 1 at s2
\revthree{
 $^\circ$CFF} &$4s(1-s)$
                  &  1 &   & ~-10.6163 & ~-9.8441 & 2.9722~ & 0.50834~ & none \\ % \qquad ~1.59617 & Phidag = 1 at s2
%%%%%%%%%%%
 $^{\circ}s$ &$80s(1-s)$ 
                  &  1 &   & ~~12.9242 & ~31.273~ & 1.4622~ & 4.0079~~ & 0.8384 \\ %%0.7974 \\ % 1.59617 & Phidag = 1 at s2
 $^\circ 1/s$&$320s(1-s)$ 
                  &  1 &   & ~~33.1890 & ~10.307~ & 3.4093~ & 1.3187~~ & 0.8963 \\ %% 0.8825 \\ % 0.63844 & Phidag = 1 at s2
%\revthree{
% $^\circ$CFF} &$160s(1-s)$
%                  &  1 &   & ~~12.0680 & ~24.854~ & 8.8326~~ & 0.56255~ & 0.9246 \\ % \qquad ~1.59617 & Phidag = 1 at s2
\revthree{
 $^\circ$CFF} &$320s(1-s)$
                  &  1 &   & ~~44.4360 & ~41.749~ & 13.789~~ & 0.67936~ & 0.8611 \\ % \qquad ~1.59617 & Phidag = 1 at s2
\end{tabular}
 \caption{Values of $\lambda$, $c$, $\alpha$, $\gamma$, and $\delta_0$ of the $n$-th mode for the
 basic magnetic field $\overline{B}$ and flow $\overline{U}$ in the spherical model $\beta = s/(1-s^2)$.
 \revthree{The CFF field $\overline{B}$ is given as $(3/2) \cos{\{ \pi(3/2 - 50 s/19) \} } + 2$.} 
 Cases indicated by $^\circ$ are evaluated with the routine bvp4c and the modified outer boundary condition.}
\label{table:cases_spheres}
  \end{center}
\end{table}
%%%%% NEW %%%%%%%%%%%%%%%%%%%%%%%%%%%%

%
%
For the Malkus field ($\overline{B} = s$) and no mean flow $\overline{U}$,
 we let $x = 1-s^2$ and $\Phi(x) = x y(x)$ to rewrite the ODE (\ref{eq:g1_ode_sphere}) as 
 \begin{equation}
    x(1-x) \frac{d^2 y}{dx^2} + (2-3x) \frac{dy}{dx} + (\lambda^2 -1) y =0
 \end{equation}
 where $\lambda^2 = - c/4$.
This is a hypergeometric equation, which has a solution
\begin{equation}
  \Phi (s) = (1-s^2) F(1+\lambda, 1-\lambda ; 2; 1-s^2) ,
       \quad \mbox{and\ } \quad
  \Phi^\dag  
  = \frac{\Phi}{1-s^2}  ,   \label{eq:phi1_sol_malkus}
\end{equation}
 where $F$ denotes the hypergeometric function \citep[e.g.][]{AS65}.
The eigenvalue $\lambda$ is determined by the condition $\Phi=0$ at $s=\eta$.
The adjoint solution is related to the axial electrical current generated at this order
 as $-D^2 \Phi =  -c \Phi s\beta/\overline{B}^2 = -c \Phi^\dag$,
 implying the current is nonzero at $s=1$.

\begin{figure}
 \centerline{
  \includegraphics[height=32mm]{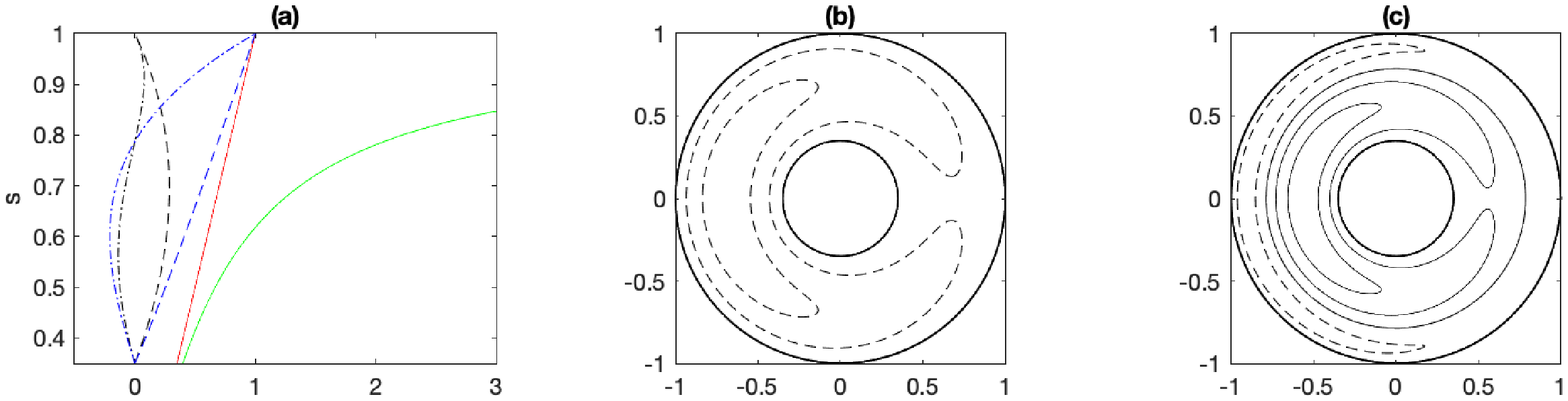} }
\caption{Spherical case for the Malkus field $\overline{B}=s$ and $\overline{U}=0$. 
 (a) Profiles of $\overline{B}$ [red solid curve], $\beta$ [green solid],
 $\Phi$ for $n=1$ [black dashed] and $n=2$ [black dashed-dotted],
 and $\Phi^\dag$ for $n=1$ [blue dashed] and $n=2$ [blue dashed-dotted].
 Streamfunctions $\psi$ of the single soliton solution for (b) $n=1$ and (c) $n=2$,
 provided $\epsilon = 0.1$.
 The dashed (solid) contour lines represent its negative (positive) value, i.e. clockwise (anti-clockwise).}
%% three ellipses:\protect\\
%%    ---$\!$---,
%%    $b/a=1$; $\cdots$\,$\cdots$, $b/a=1.5$.
 \label{fig:sphere_Malkus}
\end{figure}

Figure \ref{fig:sphere_Malkus} shows the solutions in the Malkus case. 
\revone{Figure \ref{fig:sphere_Malkus}(a)} shows profiles of $\overline{B}(s)$, the topography $\beta$, 
 the eigenfunctions $\Phi$ for $n = 1$ and $2$, 
 and their adjoint eigenfunctions $\Phi^\dag$ (\ref{eq:phi1_sol_malkus}). 
This yields
 $\alpha \approx -12.85$ and $\gamma \approx 0.87$
 for $n=1$; the nonlinear effect is more significant than the dispersive one.
\revone{Figure \ref{fig:sphere_Malkus}(b)} illustrates a single soliton solution
 (\ref{eq:single-soliton-psi_sphere}) of $\psi$ for $n=1$. 
If the amplitude $\epsilon$ is too large, neglected higher order terms will be significant;
 if $\epsilon$ is too small the azimuthal scale of the solitary wave is too large to fit in,  
 so we choose $\epsilon = 0.1$ as a reasonable compromise.
% the basic state $\psi_0$ is excluded
%
The streamfunction $\psi$ is negative,
 indicating a clockwise solitary eddy.
The retrogradely propagating vortex $\psi_1$ is slightly more concentrated at the outer shell
 than the magnetic potential $g_1$ (not shown).
As $c < 0$ and $\gamma > 0$,
 the dispersion term reduces the retrograde propagation speed.
We note that 
 a clockwise vortex is observed
 in Earth's core \citep{PJ08} and geodynamo simulations \citep{SJNF17}:
 its implications are discussed in the final section.

The same basic states admit high-$n$ modes with more isolated structure
 to have the KdV equations with nonzero $\alpha$ and $\gamma$ (Table \ref{table:cases_spheres}). 
The speed $|c|$ increases with $n$, confirming the dispersivity of the wave. 
The eigenfunction $\Phi$ for $n=2$
 is negative at small $s$,
 and then turns positive when $s \gtrsim 0.787$ (dashed-dotted curve in \revone{figure \ref{fig:sphere_Malkus}a}),
 so the eddy is clockwise in the outer region and anticlockwise in the inner region (\revone{figure \ref{fig:sphere_Malkus}c}).

%%%%%%%%%%%%%%%%%%%%%%%%%%%%

%%%%%%%%%%%%%%%%%%%%%%%%%%%%

%
%
We next consider the basic field given by the wire field, $\overline{B} = 1/s$,
 whilst $\overline{U} = 0$.
By using $\Phi(x) = x e^{\lambda x} y(x)$, (\ref{eq:g1_ode_sphere})
 may be reduced to a confluent Heun equation
\begin{equation}
    x(1-x) \frac{d^2 y}{dx^2} 
    + \{2 + (2 \lambda -3)x -2 \lambda x^2 \} \frac{dy}{dx} 
    + \{ (\lambda^2 +2 \lambda -1) - (\lambda^2+3 \lambda)x\}y =0 .
 \end{equation}
The solution regular at $s=1$ corresponding to the eigenvalue $\lambda$ is
\begin{equation}
 \Phi = (1-s^2) e^{\lambda(1-s^2)} H_\tx{c}(q_\tx{c}, \alpha_\tx{c}, \gamma_\tx{c},\delta_\tx{c},\epsilon_\tx{c}; 1-s^2),
    \quad \mbox{and\ } \quad
 \Phi^\dag 
  = \frac{s^4}{1-s^2} \Phi ,
  \label{eq:phi1_sol_wire}
 \end{equation}
 where $H_\tx{c}$ represents the confluent Heun function with  
 the accessory parameter $q_\tx{c} =\lambda^2 +2 \lambda - 1$ and exponent parameters
 $\alpha_\tx{c} = \lambda^2 + 3 \lambda, \gamma_\tx{c} = 2, \delta_\tx{c} = 1$ and $\epsilon_\tx{c} = 2 \lambda$ \citep{OLBC10}. 
\revone{
This case admits a simple form of the coefficients (\ref{eq:g2_KdV_sphere}) such that 
\begin{eqnarray}
 &&\alpha_0  
 = -4 \lambda^2  \int^{1-\eta^2}_0 
   x (2x+1) (1-x)^2  e^{3\lambda x} H_\tx{c}^3  \,dx  , \quad  
 \gamma_0
  =  \frac{1}{2} \int^{1-\eta^2}_{0} \frac{x^2}{1-x} e^{2\lambda x}  H_\tx{c}^2    \, dx
 ,  \nonumber  \\
&&\quad \textrm{and} \quad 
 \delta_0 
  =  \frac{1}{2} \int^{1-\eta^2}_{0} x(1-x)^2 e^{2\lambda x}  H_\tx{c}^2  \,dx  .  
\end{eqnarray}
}
To evaluate the function we use the algorithm of \citet{Mot18} below.

Figure \ref{fig:sphere_invB}(a) gives profiles of the basic state 
 and eigenfunctions.
The figure shows that $\Phi$ for $n=1$ 
 has a peak nearer the outer boundary, compared with that for the Malkus field;  
 it is still propagating retrogradely and is dispersive. 
This case yields
 $\alpha \approx -36.9$ and $\gamma \approx 1.25$
 for $n=1$ and with $\epsilon=0.1$ the soliton is a more compact, clockwise eddy (\revone{figure \ref{fig:sphere_invB}b}).
Analysis of the individual terms of the coefficient $\alpha_0$ in (\ref{eq:g2_KdV_sphere})
 implies that the presence of high order derivatives is favourable for nonlinear effects.
For $n=2$, 
 dispersive effects are enhanced compared to nonlinear ones. 
The solitary eddy is clockwise in the outer region when $s \gtrsim 0.894$
 and anticlockwise in the inner region (\revone{figure \ref{fig:sphere_invB}c}).

%%%%%%%%%%%%%%%%%%%%%%%
\begin{figure}
 \centerline{
  \includegraphics[height=32mm]{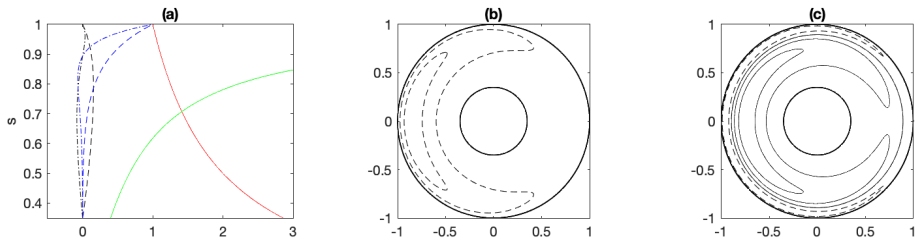} }
\caption{Spherical case for the wire field $\overline{B}=1/s$ and $\overline{U}=0$.
 (a) Profiles of $\overline{B}$ [red solid curve], $\beta$ [green solid],
 $\Phi$ for $n=1$ [black dashed] and $n=2$ [black dashed-dotted],
 and $\Phi^\dag$ for $n=1$ [blue dashed] and $n=2$ [blue dashed-dotted].
 Streamfunctions $\psi$ of the single soliton solution for (b) $n=1$ and (c) $n=2$,
 provided $\epsilon = 0.1$.}
\label{fig:sphere_invB}
\end{figure}
%%%%%%%%%%%%%%%%%%%%%%%

%
%
To explore more general cases
 we implement the Matlab routine bvp4c to solve the eigenvalue problems.
We retain the boundary condition $\Phi = 0$ at $s = \eta = 0.35$,
 but use the modified condition $\Phi + (1-s) {D\Phi} = 0$ close to the outer boundary $s=0.99999$
 to avoid the numerical issue arising from singularities when $s \rightarrow 1$. 
We also impose a normalising condition $D\Phi$ at the inner boundary: 
% the values for the Malkus field and the wire field are given
% by (\ref{eq:phi1_sol_malkus}) and (\ref{eq:phi1_sol_wire}), respectively.
\revthree{
 the values for the Malkus field and the CFF field are given by (\ref{eq:phi1_sol_malkus}),
 whereas the one for the wire field is by (\ref{eq:phi1_sol_wire}).
}
The number of gridpoints in $s$ is $500$ in all cases.
Given the obtained $c$, the same routine is adopted
 to solve the boundary value problems for $\Phi^\dag$.
For consistency with the earlier cases we set $\Phi^\dag = 1$ at the outer boundary.
The codes are benchmarked with the exact solutions.
With modified boundary condition,
 our computational results match the expected eigenvalues $\lambda = \sqrt{|c|}/2$
 and eigenfunctions $\Phi$ for $1 \le n \le 3$
 with errors less than 0.01 \% and 0.2 \%, respectively.

\begin{figure}
 \centerline{
  \includegraphics[height=32mm]{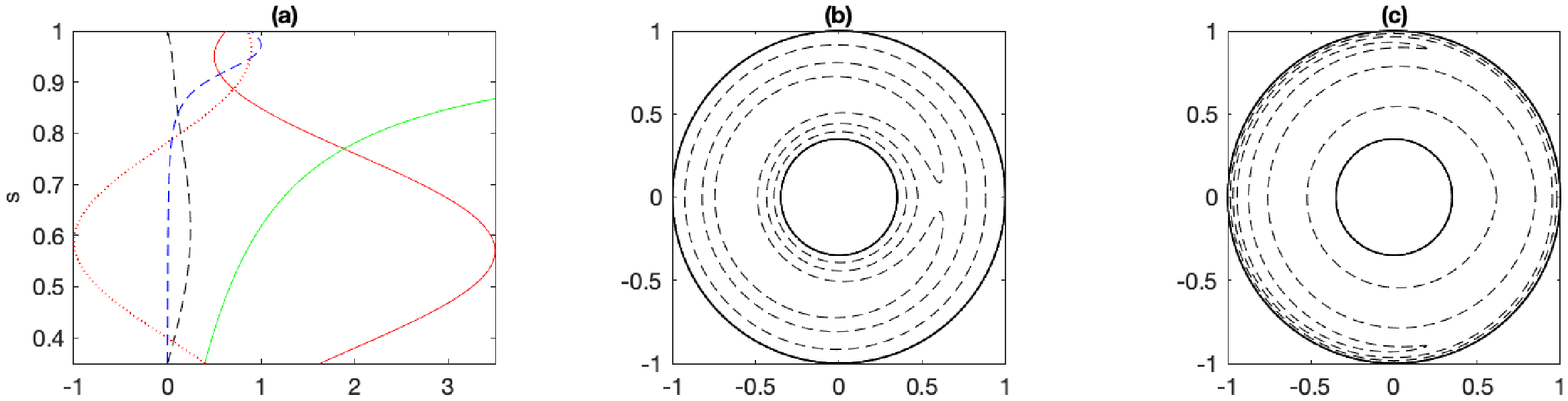} }
\caption{\revthree{Spherical case for the CFF field $\overline{B}=(3/2)\cos{\{ \pi(3/2 - 50s/19) \} }+2$,
 $\overline{U}=0$, and $n=1$.
 (a) Profiles of $\overline{B}$ [red solid curve], $\beta$ [green solid],
 $\overline{J}$ [red dotted; normalised for visualisation], $\Phi$ [black dashed], and $\Phi^\dag$ [blue dashed].
 (b) Magnetic potential $g_1$ of the single soliton solution, where the basic state $g_0$ is excluded to help visualisation.
 (c) Streamfunctions $\psi$ of the solution, provided $\epsilon = 0.1$.}
}
%
% \caption{Spherical case for the trigonometric field $\overline{B}= (3/2) \cos{\{ \pi(3/2 - 100 s/38) \} } + 2$, the basic flow $\overline{U} = 80s(1-s)$,
% and $n=1$.
% (a) Profiles of $\overline{B}$ [red solid curve], $\beta$ [green solid],
% $\overline{U}/10$ [{blue solid}; normalised for visualisation],
% and the deviation $\overline{U}/s - c$ [blue dotted]. 
% (b) Profiles of $\Phi$ [black dashed], $\Phi^\dag$ [blue dashed], and $D \Phi$ \revone{[black dotted]}.
% (c) Streamfunction $\psi_1$ of the single soliton solution,
% where the basic state $\psi_0$ is excluded to help visualisation.}
\label{fig:sphere_CFF}
\end{figure}

\revthree{
Now the third basic field, $\overline{B} = (3/2) \cos{\{ \pi(3/2 - 50 s/19) \} } + 2$, is examined.
Figure \ref{fig:sphere_CFF}(a) depicts the basic state, the eigenfunctions for $n=1$, 
 and additionally $\overline{J}$ (represented by the red dotted curve).
It is nonzero except at $s \approx 0.40$ and $0.78$
 and is negatively peaked at $s \approx 0.59$. 
The eigenvalues $c$ do not differ from those in the Malkus case very much (Table \ref{table:cases_spheres}). 
For $n=1$,
 $\Phi$ has a peak at $s \approx 0.61$ (blue dashed curve), as so does the basic field. 
This case gives $\alpha \approx -11.5$ and $\gamma \approx 2.85$. 
Indeed the term including $\overline{J}$ dominates over the ODE (\ref{eq:g1_ode_sphere})
 and also over $\alpha_0$ (\ref{eq:g2_KdV_sphere});
 if the term $\Phi^2 D(\overline{J}/\beta)$ were absent, $\alpha$ would become $\approx 1.68$.
Figure \ref{fig:sphere_CFF}(b) illustrates the magnetic potential $g_1$ (\ref{eq:single-soliton-g_sphere}),
 where the basic state is excluded for visualisation, 
It is clockwise and centred at the $s \approx 0.61$.
Similarly the streamfunction $\psi$ (\ref{eq:single-soliton-psi_sphere}) is displayed
 in figure \ref{fig:sphere_CFF}(c): 
 now the distinction from the magnetic component is evident. 
The solitary eddy is more confined nearer the outer boundary,
 as $\overline{B} D^2 \Phi - \overline{J}\Phi$ in (\ref{eq:single-soliton-psi_sphere}) becomes significant
 only when $s \gtrsim 0.8$ (not shown).
}

%%%%%%%%%%%%%%%%%%%%%%%%%%%%%%
%%%%%%%%%%%%%%%%%%%%%%%%%%%%%%

%
%
Including a basic flow $\overline{U} = s$
 is equivalent to the addition of solid body rotation. 
Therefore it affects the speed $c$ of propagation of the mode,
 whilst leaving its other properties unchanged (Table \ref{table:cases_spheres}). 
For a more realistic flow, $\overline{U} = 4 s(1-s)$, with the Malkus field,
 the structures of $\Phi$ and $\Phi^\dag$ are not drastically altered
 (leading to $\delta_0 \approx 0.049$).
The dominance of the nonlinearity over the dispersion, $|\alpha/\gamma|$, is however weakened.
The presence of the same basic flow in the wire field case also exhibits this property.

\begin{figure}
 \centerline{
  \includegraphics[height=32mm]{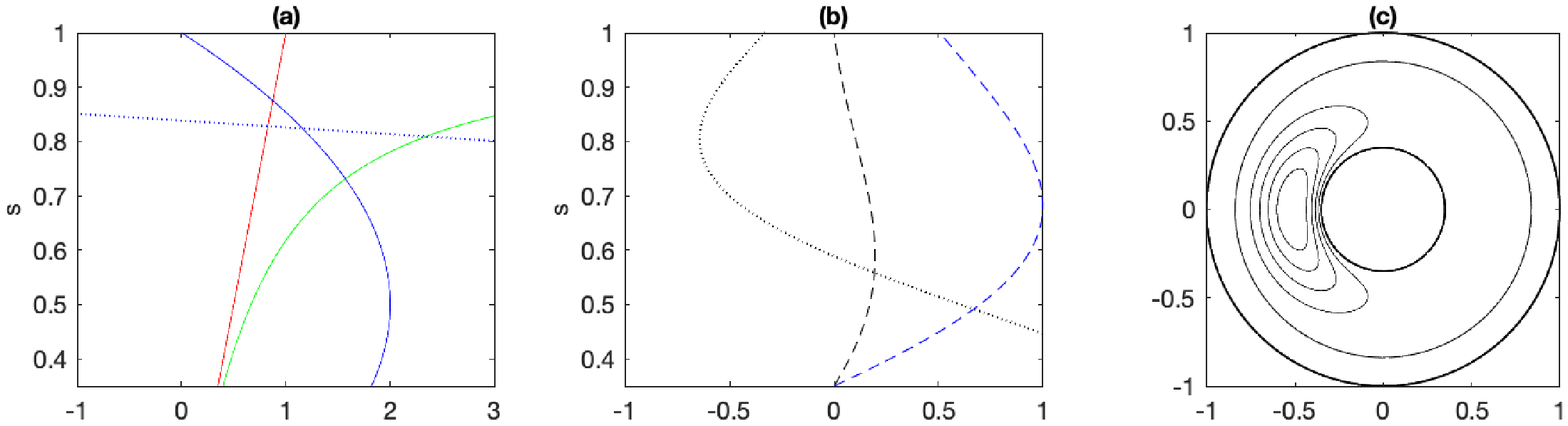} }
 \caption{Spherical case for the Malkus field $\overline{B}=s$, the basic flow $\overline{U} = 80s(1-s)$,
 and $n=1$.
 (a) Profiles of $\overline{B}$ [red solid curve], $\beta$ [green solid],
 $\overline{U}/10$ [{blue solid}; scaled for visualisation],
 and the deviation $\overline{U}/s - c$ [blue dotted]. 
 (b) Profiles of $\Phi$ [black dashed], $\Phi^\dag$ [blue dashed], and $D \Phi$ \revone{[black dotted]}.
 (c) Streamfunction $\psi_1$ of the single soliton solution,
 where the basic state $\psi_0$ is excluded to help visualisation.}
\label{fig:sphere_Malkus_shear}
\end{figure}

Finally, we comment on the behaviour of solutions in the vicinity of the point $s$ 
 at which $\overline{U}/s$ equals $c$,
 the location of a critical layer for the hydrodynamic Rossby wave soliton \citep[e.g.][]{R77}. 
We impose a fast mean zonal flow, $\overline{U} = 80 s(1-s)$, in the Malkus field case; 
 figure \ref{fig:sphere_Malkus_shear}(a) shows the basic state
 and additionally the deviation from the wave speed, $\overline{U}/s - c$ (blue {dotted} curve). 
The curve shows this case has such a critical point at $s \approx 0.838$.  %%0.797$}.
Nevertheless the impact is hardly seen in the eigenfunctions $\Phi$ and $\Phi^\dag$:
 there are no discontinuities in the derivative $D\Phi$ (\revone{figure \ref{fig:sphere_Malkus_shear}b})
 and hence in the solitary wave solutions (\revone{figure \ref{fig:sphere_Malkus_shear}c}). 
This remains true for the wire field case with $\overline{U}=320s(1-s)$.

%%%%%%%%%%%%%%%%%%%%%%%%%%%%%%
%%%%%%%%%%%%%%%%%%%%%%%%%%%%%%

\section{Concluding remarks}

In this paper we have performed a weakly nonlinear analysis of magnetostrophic waves
 in QG spherical models with azimuthal  magnetic fields and flows. 
The model we considered is an annulus model \citep{Bus76,CFF14} of the form
 utilised by \citet{H66} for linear magnetic Rossby (MR) waves.
We found that 
 the evolution of the long-wavelength, slow-MR waves in the spherical shells obeyed the KdV equation,
 whether the toroidal magnetic field and/or the zonal flow were sheared or not. 
The model we consider here is formally valid for cases where the azimuthal lengthscale is
 much longer than that in radius; the most obvious application of which is for thin spherical shells. 
For thicker spherical shells like those representative of Earth's fluid outer core,
 the ratio of these lengthscales is of the order ten.
For thinner shells relevant to other astrophysical objects one might expect the asymptotic procedure
 to give a better approximation to the true behaviour.
We find that solutions may take the form of a single soliton solution (for $n=1$)
 which is a clockwise, solitary eddy when basic state magnetic field is %either
 \revthree{any of} a Malkus field ($\overline{B} \propto s$), %or
 a magnetic wire field ($\overline{B} \propto 1/s$)\revthree{, and
 a CFF field (comprising of a trigonometric function)}.
In addition to these steadily progressing single-solitons we also find $N$-soliton solutions;
 as these satisfy the KdV equation we know that these may have peculiar interactions
 including a phase shift after a collision and FPU recurrence \citep[e.g.][]{DJ89}.

We conclude by noting that inversion of the geomagnetic secular variation appears
 to detect an anticyclonic gyre in Earth's core \citep{PJ08,GJF15,BHFMG18}; 
 it is off-centred with respect to the rotation axis
 and is believed to have existed for more than a hundred years.
Moreover, DNS of dynamos driven by convection in rapidly-rotating spherical shells
 have exhibited the emergence of a large vortex
 which circulated clockwise and modulated very slowly \citep{SJNF17}; in these simulations
 the averaged toroidal magnetic field tended to strengthen beneath the outer boundary. 
Our solution tentatively supports the idea that such an isolated single eddy should persist,
 while drifting on MC timescales of %$\mathcal{O}(10^{2\textrm{-}3})$ years. 
 \revthree{$\mathcal{O}(10^{2\textrm{-}4})$ years}.
 The long wave can be initiated through instabilities %, for example,
 due to differentially rotating flows \citep{SAetal08}, % ,NJSRG10}, 
 due to thermally insulating boundaries \citep{HTS14}, % S02,
 and due to the magnetic diffusivity \citep{RL79,ZLS03}. 
The steadily drifting feature of the solitons should of course be altered during the long-term evolution
 when dissipation plays a role in the dynamics. 
\revthree{The presence of dissipation may also alter the eigenfunction \citep{CFF14}
  and thus the detailed morphology of the soliton too.}

\revthree{
We note an alternative to account for the eccentric gyre is
 a flow induced by, for example, the coupling with the rocky mantle and the solid inner core,
 as DNS by \citet{AFF13} had demonstrated.
%Incorporating with a horizontal 
% may enable to explain the eccentricity and the location of the gyre, 
%% as observed in the core flow inversion:
% as DNS by \citet{AFF13} had demonstrated.}
The issue ends up in a debate which has lasted for decades: 
 does the geomagnetic westward drift represent the advection due to a large scale fluid motion 
 \citep{BFGN50} or hydromagnetic wave motion \citep{H66}.}
We shall investigate these issues further, as well as the role of critical layers,
 by solving initial value problems in a future study.

\begin{acknowledgments}
\section*{Acknowledgments}
%Acknowledgements should be included at the end of the paper, before the References section or any appendicies, and should be a separate paragraph without a heading. Several anonymous individuals are thanked for contributions to these instructions.

The authors are grateful to
 Andrew Soward, Anna Kalogirou, Adrian Barker, 
 and Yoshi-Yuki Hayashi for discussion and comments.
%Comments by ** helped to improve the manuscript.
%
K.~H. was supported by the Japan Science and Technology under 
 the Program to Supporting Research Activities of Female Researchers.
\end{acknowledgments}

%%%%%%% referencese %%%%%%
\bibliographystyle{jfm}
%%\bibliography{jfm_HTJ}

\end{document}